\documentclass[prl,letterpaper,aps,floatfix,twocolumn,superscriptaddress]{revtex4-1}
\usepackage{graphicx}
\usepackage{amsmath}
\newcommand{\beq}{\begin{equation}}
\newcommand{\eeq}{\end{equation}}
\newcommand{\bk}{{{\bf{k}}}}

\newcommand{\beqa}{\begin{eqnarray}}
\newcommand{\eeqa}{\end{eqnarray}}

\newcommand{\dg}{{\dag}}
\newcommand{\pdg}{{\vphantom\dag}}
\newcommand{\btau}{{\boldsymbol \tau}}
\newcommand{\bsigma}{{\boldsymbol \sigma}}

\begin{document}
\title{Weyl Semimetal in a Topological Insulator Multilayer}
\author{A.A. Burkov}
\affiliation{Department of Physics and Astronomy, University of Waterloo, Waterloo, Ontario 
N2L 3G1, Canada}
\affiliation{Kavli Institute for Theoretical Physics, University of California, Santa Barbara, CA 93106, USA}
\author{Leon Balents}
\affiliation{Kavli Institute for Theoretical Physics, University of California, Santa Barbara, CA 93106, USA} 
\date{\today}
\begin{abstract}
  We propose a simple realization of the three-dimensional (3D) Weyl
  semimetal phase, utilizing a multilayer structure, composed of
  identical thin films of a magnetically-doped 3D topological
  insulator (TI), separated by ordinary-insulator spacer layers.  We
  show that the phase diagram of this system contains a Weyl semimetal
  phase of the simplest possible kind, with only two Dirac nodes of
  opposite chirality, separated in momentum space, in its
  bandstructure.  This Weyl semimetal has a finite anomalous Hall conductivity, 
  chiral edge states, and occurs as an intermediate phase between an ordinary insulator and a 3D quantum
  anomalous Hall insulator. We find that the Weyl semimetal has a
  nonzero DC conductivity at zero temperature, but Drude weight vanishing as  $T^2$, and is thus an unusual
  metallic phase, characterized by a finite anomalous Hall
  conductivity and topologically-protected edge states.
\end{abstract}
\maketitle
The recent discovery of time-reversal (TR) invariant topological
insulators (TI)~\cite{Kane05} has led to a surge of
interest in topological properties of the electronic bandstructure of
crystalline materials.  TIs exhibit a bulk gap, but gapless
surface states, whose gaplessness is protected by topology.
Remarkably, recent work has demonstrated that such a surface-bulk
correspondence can also obtain even when the bulk is gapless, by
virtue of point touchings of non-degenerate conduction and valence
bands~\cite{Vishwanath11}.  Such accidental point touchings have been
known to exist since the earliest days of the theory of
solids~\cite{Herring37}, but their topological properties have only
been appreciated much more
recently~\cite{Vishwanath11,Volovik,Murakami07,Ran11}, and
concrete materials, where they may be found, have been
proposed~\cite{Vishwanath11,Ran11}.  Non-trivial and robust band
touching requires either broken TR or inversion
(I)~\cite{Herring37,Murakami07,Vishwanath11}, in which case the
touching points acquire topological character and thus give rise to
stable phases of matter.  The bandstructure near these points can be
described by a massless two-component Dirac or Weyl Hamiltonian: 
\beq
\label{eq:1}
{\cal H} = \pm v_F \bsigma \cdot \bk, 
\eeq 
where $\bk$ is the crystal momentum in the first Brillouin zone (BZ), 
expanded near the band-touching point, $\bsigma$ is the
triplet of Pauli matrices and the sign in front corresponds to two
different possible chiralities, characterizing the point.  Such Weyl
fermions have been studied extensively in high-energy physics, in particular as a description of neutrinos~\cite{Volovik},
and may be viewed as topological defects (hedgehogs) in momentum
space~\cite{Volovik}.  Any perturbation of Eq.(\ref{eq:1}) only shifts
the degeneracy point in energy or momentum, but does not remove it: an
isolated Weyl fermion in this sense possesses an absolute topological
stability (this is in contrast to 2D massless Dirac fermions 
in graphene, where inversion symmetry of the honeycomb lattice is essential for 
their stability).  
Very general considerations show that Dirac degeneracy points can only
occur in pairs of opposite chirality~\cite{Nielsen81} and can thus be
eliminated by pairwise annihilation.  When the TR or I symmetry is
broken, however, the Weyl fermions are separated in momentum space and
thus, assuming translational symmetry remains intact, are still
topologically stable.

Ref.~\cite{Vishwanath11} has proposed a possible realization of a Weyl
semimetal with 24 Dirac nodes in iridium pyrochlores, which are strongly-correlated
magnetic materials (a different scenario
for this material was proposed in Ref.~\cite{Pesin10}). 
The purpose of this work is to propose a much simpler realization of
the Weyl semimetal, not relying on strong correlations in a rather
complex material.  The Weyl semimetal we propose also possesses only
two Dirac nodes, the smallest possible number, and thus is in a sense
the most elemental realization of this phase of matter.

\begin{figure}[t]
\includegraphics[width=6cm]{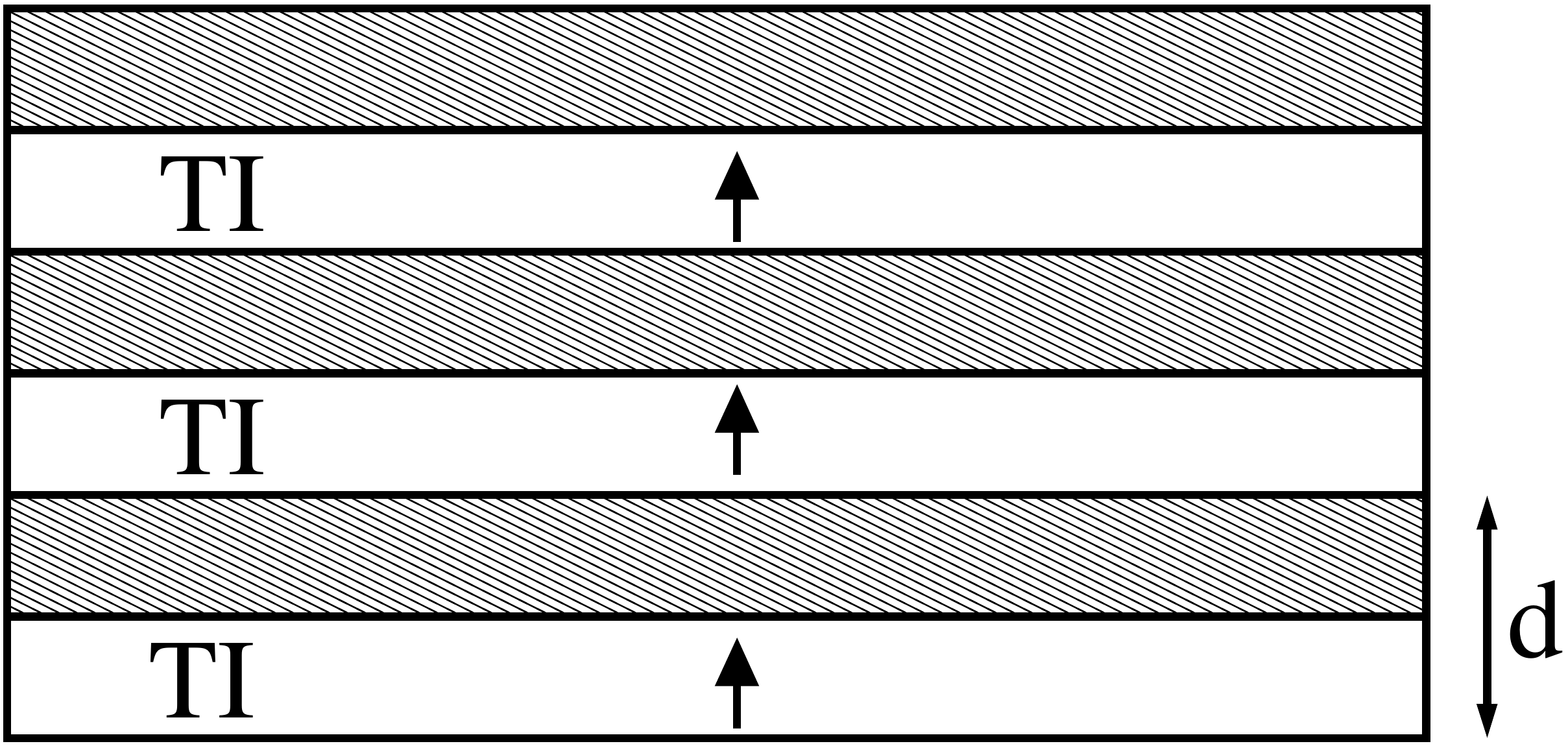}
\caption{Schematic drawing of the proposed multilayer structure. Unhashed layers are the TI layers, while 
hashed layers are the ordinary insulator spacers. Arrow in each TI layer shows the magnetization direction.
Only three periods of the superlattice are shown in the figure, 20-30 unit cells can perhaps be grown realistically.} 
\label{fig:1}
\end{figure}

The material we propose is a multilayer heterostructure, consisting of
alternating layers of a 3D TI material, such as $\textrm{Bi}_2
\textrm{Se}_3$, and an ordinary insulator, which serves as spacer
material between the neighboring TI layers, as shown in
Fig.~\ref{fig:1}.  The ability to grow ultrathin high-quality films of
$\textrm{Bi}_2 \textrm{Se}_3$ has been clearly demonstrated in recent
experiments~\cite{ZhangG09}. It is thus quite realistic to
expect that a multilayer heterostructure, consisting essentially of a
stack of such thin films, can be fabricated using available
technology.  The Hamiltonian, describing this heterostructure, can be
written as: 
\beqa
\label{eq:2}
H&=&\sum_{\bk_{\perp}, ij} \left[ v_F \tau^z (\hat z \times \bsigma) \cdot \bk_{\perp} \delta_{i,j} + m \sigma^z \delta_{i,j}
+ \Delta_S \tau^x \delta_{i,j} \right. \nonumber \\ 
&+& \left.\frac{1}{2} \Delta_D \tau^+ \delta_{j, i+1} + \frac{1}{2} \Delta_D \tau^- \delta_{j, i-1} \right] c^\dg_{\bk_{\perp} i} c^\pdg_{\bk_{\perp} j}. 
\eeqa
The first term in Eq.(\ref{eq:2}) describes the two (top and bottom) surface states of an individual TI layer. 
We assume for simplicity that a TI material with a single two-dimensional (2D) Dirac node per surface BZ is employed. 
$v_F$ is the Fermi velocity, characterizing the surface Dirac fermion, which we take to be the same on the top and 
bottom surface of each layer. $\bk_{\perp}$ is the momentum in the 2D surface BZ (we use $\hbar = 1$ units), 
$\bsigma$ is the triplet of Pauli matrices acting on the real spin degree of freedom and $\btau$ are Pauli matrices acting 
on the {\em which surface} pseudospin degree of freedom. The indices $i,j$ label distinct TI layers. 
The second term describes exchange spin splitting of the surface states, which can be induced, for example, by doping 
each TI layer with magnetic impurities, as has been recently demonstrated experimentally~\cite{Chen10}.
The remaining terms in Eq.(\ref{eq:2}) describe tunneling between top and bottom surfaces within the same TI layer (the term, proportional 
to $\Delta_S$), and between top and bottom surfaces of neighboring TI layers (terms, proportional to $\Delta_D$).  
Longer-range tunneling is assumed to be negligible. We will regard $m$ and $\Delta_{S,D}$ as tunable parameters and study 
the phase diagram of Eq.(\ref{eq:2}) as a function of these parameters. 

Let us initially assume that the spin splitting is absent, i.e. set $m = 0$. 
Diagonalizing Eq.(\ref{eq:2}) one finds the following band dispersion:
\beq
\label{eq:3}
\epsilon_{\bk \pm}^2 = v_F^2 (k_x^2 + k_y^2) + \Delta^2(k_z),
\eeq
where $\Delta(k_z) = \sqrt{\Delta_S^2 + \Delta_D^2 + 2 \Delta_S \Delta_D \cos(k_z d)}$, 
and $d$ is the superlattice period (i.e. TI layer plus spacer layer thickness) in the growth ($z$) direction. 
This bandstructure is fully gapped when $|\Delta_S| \neq |\Delta_D|$, but contains Dirac nodes 
 when $\Delta_S/\Delta_D = \pm 1$. The nodes are located at $k_z = \pi/d$ when $\Delta_S/\Delta_D = 1$ and 
 at $k_z = 0$ when $\Delta_S/\Delta_D = - 1$ ($k_x = k_y = 0$ always). While both cases are possible, we will assume the former for concreteness and will take both tunneling matrix elements to be positive (this choice does not affect any of our results). 
Expanding the band dispersion near the Dirac point at $k_x = k_y = 0, k_z = \pi/d$ to leading order in the momentum one obtains: 
\beq
\label{eq:4}
\epsilon_{\bk \pm}^2 = v_F^2 (k_x^2 + k_y ^2) + \tilde v_F^2 k_z^2, 
\eeq
where $\tilde v_F = d \sqrt{\Delta_S \Delta_D}$.  
The momentum-space Hamiltonian near the Dirac node has the form:
\beq
\label{eq:5} {\cal H}(\bk) = v_F \tau^z (\hat z \times \sigma) \cdot
\bk + \tilde v_F \tau^y k_z, \eeq which can be brought to a
block-diagonal form, explicitly revealing a pair of two-component Weyl
fermions with opposite chirality, by a $\pi/2$ rotation around the
$x$-axis in the pseudospin space. 
Alternatively, in total this is a conventional 4-component massless Dirac fermion.  
\begin{figure}[t]
  \centering
  \includegraphics[width=7cm]{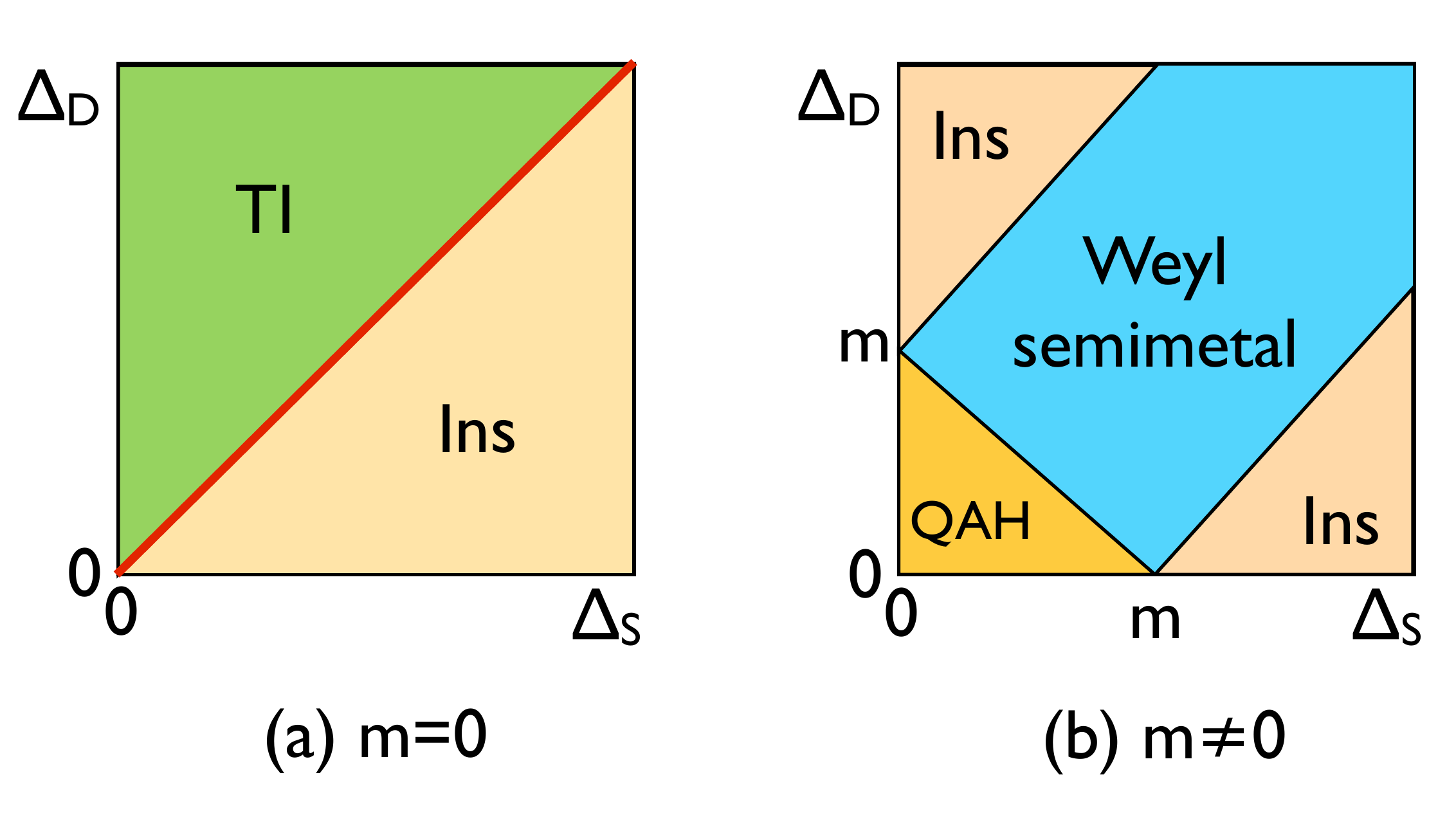}
  \caption{(Color online) Phase diagrams for (a) $m=0$ and (b) $m \neq 0$.  In (a),
    the red line represents the phase boundary between topological
    insulator (TI) and ordinary insulator (Ins).  In (b), due to
    TR symmetry breaking, the distinction between topological
    and ordinary insulators is moot, so the TI in (a) has been
    converted to Ins.  QAH denotes the quantum anomalous Hall phase. }
  \label{fig:pd}
\end{figure}
As discussed above, since the two Weyl fermions are located at the
same point in momentum space, they are topologically unstable. Any
perturbation, for example any deviation of the ratio
$\Delta_S/\Delta_D$ from unity, immediately eliminates the degenerate
Dirac node and produces a fully gapped spectrum.  With $m=0$, the
massless Dirac point can be understood~\cite{Murakami07}\ as a critical
point between topological ($\Delta_D>\Delta_S$) and ordinary
($\Delta_D<\Delta_S$) insulators with both inversion and time-reversal
symmetry preserved (see Fig.~2a). 
To produce a topologically stable phase with 3D
Dirac nodes, the nodes have to be separated in momentum space.  As
mentioned above, this can generally be accomplished by breaking either
TR or I symmetries and there are in principle many ways to do
this. Here we will focus on one particular way, which is perhaps the
simplest from the point of view of a practical realization.  Namely,
as already mentioned above, we will assume that each TI layer is doped
with magnetic impurities, producing a ferromagnetically-ordered state
within each layer, with magnetization along the growth direction of
the heterostructure.  This leads to spin splitting of the surface
states of magnitude $m$, described by the second term in
Eq.(\ref{eq:2}).  The band dispersion is now given by: 
\beq
\label{eq:6}
\epsilon_{\bk \pm}^2 = v_F^2 (k_x^2 + k_y^2) + \left[ m \pm \Delta(k_z) \right]^2. 
\eeq 
This dispersion has two nondegenerate Dirac nodes, separated along the $z$-axis in momentum space, with locations 
given by $k_z = \pi/d \pm k_0$, where: 
\beq
\label{eq:7}
k_0 = \frac{1}{d} \arccos\left[1- (m^2 - (\Delta_S - \Delta_D)^2)/ 2 \Delta_S \Delta_D \right]. 
\eeq
The nodes exist as long as:
\beq
\label{eq:8}
m^2_{c1} = (\Delta_S - \Delta_D)^2 < m^2 < m^2_{c2} = (\Delta_S + \Delta_D)^2. 
\eeq 
Thus, as discussed above, splitting of the degenerate Dirac node in momentum space produces a stable Weyl semimetal {\em phase}, 
existing in a finite region of the phase diagram (Fig.~2b). 
The Weyl semimetal occurs as an intermediate phase between an ordinary insulator ($m^2 < m_{c1}^2$) and 
a 3D quantum anomalous Hall (QAH) insulator~\cite{Yu10} with quantized Hall
conductivity, equal to $e^2/h$ per TI layer ($m^2 > m_{c2}^2$).
As $m$ is increased from zero, a degenerate Dirac point at $k_x = k_y = 0, k_z = \pi/d$ appears at the lower critical 
value ($m_{c1}$) of the spin splitting.
The Dirac node is split along the $z$-axis into two nondegenerate nodes in the Weyl semimetal phase, with the splitting increasing 
monotonically with the magnitude of $m$. 
At the upper critical value ($m_{c2}$), the two nondegenerate nodes meet and annihilate at the 
center of the BZ  $k_x = k_y = k_z = 0$, giving rise to the fully gapped QAH insulator. 

We will now show that the Weyl semimetal phase we have found is characterized 
by a nonzero, but nonquantized anomalous Hall conductivity, proportional to the magnitude of the separation of the 
Dirac nodes in momentum space, thus smoothly evolving from zero in the ordinary insulator phase to $e^2/h$ per layer 
in the 3D QAH insulator phase. 

The fact that a 3D Weyl semimetal generally has a finite Hall conductivity is known, and was pointed out 
e.g. in Ref.~\cite{Volovik05}. 
For the particular realization of a Weyl semimetal, proposed in~\cite{Vishwanath11}, the Hall conductivity vanishes 
due to the cubic symmetry of the crystal structure of the pyrochlore iridate materials. 
If the cubic symmetry is broken, e.g. by applying a uniaxial pressure, as proposed in~\cite{Ran11}, the Hall conductivity 
becomes nonzero, and is proportional, in the case discussed in~\cite{Ran11}, to the applied pressure. 
In our case, there is a natural preferred axis, i.e. the growth direction of the heterostructure (which coincides 
with the magnetization direction) and the Hall conductivity in the Weyl semimetal phase is automatically nonzero. 
The simplest and most physically transparent way to obtain this result is to view the 3D bandstructure as a set of independent 2D
bandstructures at fixed $k_z$.  We begin with the momentum-space Hamiltonian: 
\beqa
\label{eq:9}
{\cal H}(\bk)&=&v_F \tau^z (\hat z \times \bsigma) \cdot \bk  + m
\sigma^z  + \hat\Delta(k_z),
\eeqa
where $\hat{\Delta} = \Delta_S \tau^x + \frac{1}{2}\left(\Delta_D
  \tau^+ e^{i k_z d} + h.c. \right)$. This is simplified by the canonical transformation: 
\begin{equation}
  \label{eq:19}
  \sigma^\pm \rightarrow \tau^z \sigma^\pm, \qquad \tau^\pm
  \rightarrow \sigma^z \tau^\pm.
\end{equation}
After this transformation, the Hamiltonian becomes:
\begin{equation}
  \label{eq:20}
  \mathcal{H}({\bf k}) = v_F k_y \sigma^x - v_F k_x \sigma^y + 
  [ m + \hat\Delta(k_z) ] \sigma^z.
\end{equation}
Now $\hat{\Delta}$ is a constant of motion, and may be replaced by
its eigenvalues $\hat\Delta(k_z)=\pm \Delta(k_z)$.  For each of these
cases, Eq.~\eqref{eq:20} gives a 2D Dirac Hamiltonian for
fixed $k_z$, with a mass $M_\pm=m\pm \Delta(k_z)$.  For $m > m_{c1}$,
$M_{-}$ vanishes when $k_z=\pi/d \pm k_0$, corresponding to the two
Dirac nodes ($M_+$ never vanishes).  

\begin{figure}[t]
\includegraphics[width=7cm]{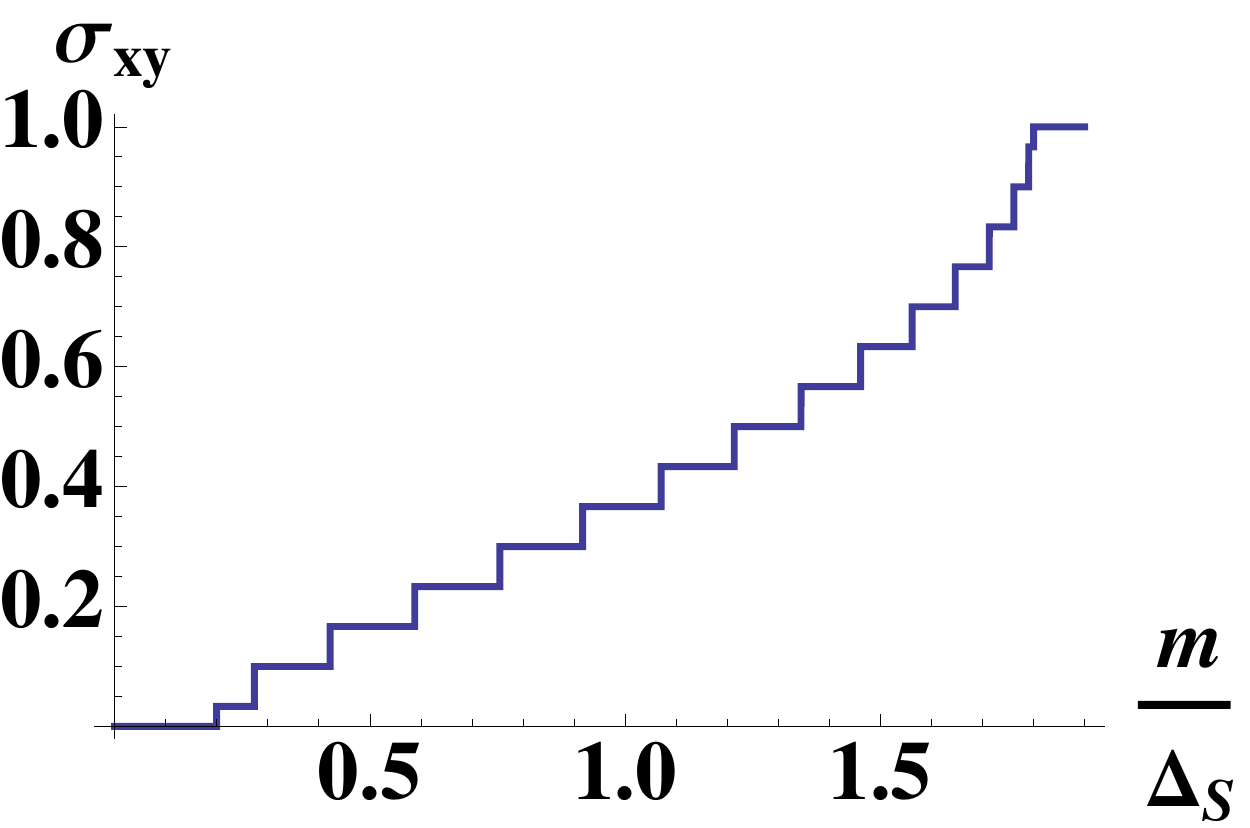}
\caption{(Color online) A plot of $\sigma_{xy}$ in units of $e^2/h d$ for a multilayer consisting of 30 TI layers. 
$\Delta_D$ is taken to be equal to $0.8 \Delta_S$. 
The Hall conductivity is zero in the ordinary insulator phase ($m < \Delta_S - \Delta_D$) and reaches the maximum value of $e^2/ h d$ 
in the QAH phase ($m > \Delta_S + \Delta_D$). 
Plateaus correspond to the wavevector $k_0$ being in an interval between the neighboring finite-size-quantized $k_z$ values.}
\label{fig:2}
\end{figure}

It is well-known that a change of sign of a 2D Dirac mass
signals a quantum Hall transition, at which the quantized 2D 
Hall conductivity $\sigma_{xy}^{2D}$ jumps by $e^2/h$~\cite{Ludwig94}.
The absolute value of the Hall conductance is not determined by the above
continuum model.  Therefore the contribution to
the total 3D Hall conductance of the states at fixed $k_z$ is equal to
$\sigma_{xy}^{2D}(k_z) = e^2/h\, [n + \Theta(k_0 - |k_z - \pi/d|)]$, where
$\Theta(x)$ is the Heaviside step function, and $n$ is an integer.
Since when $m$ vanishes TR symmetry demands that the Hall
conductivity must also vanish, we can conclude that $n=0$, and hence:
\begin{equation}
  \label{eq:21}
  \sigma_{xy} = \int_{-\pi/d}^{\pi/d} \! \frac{dk_z}{2\pi} \,
  \sigma_{xy}^{2D}(k_z) = \frac{e^2 k_0}{\pi h}.
\end{equation}
Thus the anomalous Hall conductivity in the Weyl semimetal is
proportional to the separation of the Dirac nodes in momentum space.
For a multilayer, consisting of a finite number of layers, as will be
the case in experiment, $\sigma_{xy}$ will exhibit plateaus as a
function of $m$, when $k_0$ will fall in an interval between the
neighboring quantized $k_z$ values, as shown in Fig.~\ref{fig:2}.  At
the upper critical value of $m = m_{c2}$, $2 k_0 = 2 \pi/d$, the two
Dirac nodes annihilate each other at the center of the BZ, and the
Hall conductivity reaches a quantized value per TI layer: 
\beq
\label{eq:16}
\sigma_{xy} = \frac{e^2}{d h},  
\eeq
which characterizes the $m > m_{c2}$ 3D QAH insulator phase.

The results of Ref.~\cite{Vishwanath11} imply the existence of ``Fermi
arcs'' for the Weyl semimetal, in this case for any surface except the one
normal to the $z$ axis.  In fact, this arc is nothing but the set of
edge states corresponding to the 2D integer quantum Hall states for
$\pi/d-k_0 <|k_z|< \pi/d$.  These can be explicitly found from
Eq.~\eqref{eq:20} for e.g. the
case of a surface at $y=0$, modeled by a $y$-dependent spin-splitting field $m(y)$.
Outside the sample ($y>0$) we take $m(y\rightarrow +\infty)=0$, which
realizes an ordinary insulator, while $m > m_{c1}$ for $y<0$, and consider
the eigenstates for fixed $k_x, k_z$.  There are then special surface
wavefunctions with $\hat{\Delta}(k_z) = -\Delta(k_z)$, which are eigenstates of \eqref{eq:20}:
\begin{equation}
  \label{eq:22}
  \psi_{\rm surf}(k_x, k_z; y) = e^{\int_0^y dy' [m(y')-\Delta(k_z)]/v_F} |\sigma^y=-1\rangle.
\end{equation}
For $\pi/d-k_0 <|k_z|< \pi/d$ and {\em not} otherwise, the
exponential above vanishes when $y \rightarrow \pm \infty$ and the state is normalizable and localized
to the surface with a localization length $\xi(k_z) \sim
v_F/M_-(k_z)$.  The energy of this state is simply $\epsilon_{\rm
  surf} = v_F k_x$, indeed identifying it as a chiral edge state.  As
$m$ is increased to enter the QAH insulator phase, the arc extends
across the full BZ and then the surface states can be alternatively viewed
in a Wannier basis localized in individual TI layers, i.e. as conventional
quantum Hall edge states for each layer.

While the Weyl semimetal has topologically-protected edge states
(topological protection follows from the separation of the Dirac nodes
in momentum space and from the fact the edge states are chiral), it is
actually {\em not a Hall insulator}.  At first sight, the low density of
states $g(\epsilon) \sim \epsilon^2$, seems to suggest a vanishing DC
conductivity at zero temperature~\cite{Vishwanath11}. However, a careful
calculation reveals that, in the presence of disorder but not
interactions, this is not the case. Using the standard Kubo formula
expression (or Boltzmann equation) with the Born-approximation impurity
scattering rate $1/ \tau(\epsilon) = 2 \pi \gamma g(\epsilon)$, where
$g(\epsilon) = \epsilon^2 / 2 \pi^2 v_F^3$ and $\gamma$ characterizes
the strength of the impurity potential, the optical conductivity of a
Weyl semimetal with isotropic Fermi velocity $v_F$ at temperature $T$,
is given by: \beq
\label{eq:sigmadc}
\textrm{Re} \, \sigma(\omega) \sim \frac{e^2 v_F^2}{h \gamma}
\int_{-\infty}^{\infty} d x\,\,\frac{x^4 \,\textrm{sech}^2(x)}{x^4 +
  (h^3 v_F^3 \omega/ 32 \pi^2 \gamma T^2)^2 }, \eeq where we have
restored explicit $\hbar$ for clarity.  This gives a finite DC
conductivity $\sigma_{DC} \sim e^2 v_F^2 / h \gamma$, which can be
expected to be large in a clean multilayer, but a Drude-like peak in the
optical conductivity with weight, vanishing as $T^2$.  Thus with
disorder (but neglecting interactions), the Weyl semimetal is not an
insulator, but an unusual metal, characterized by a nonzero anomalous
Hall conductivity and topologically-protected edge states.
 
In conclusion, we have proposed a simple realization of a 3D Weyl semimetal phase in a multilayer structure, composed
of a stack of thin layers of magnetically-doped 3D TI material, separated by insulating spacers. 
We have shown that this material realizes the simplest possible type of Weyl semimetal, with only two Dirac nodes, 
separated along the growth direction of the heterostructure in momentum space.
This Weyl semimetal is characterized by a nonzero anomalous Hall conductivity, proportional to the separation between 
the Dirac nodes, and by the existence of topologically stable chiral edge states. These edge states are, however, distinct from the ordinary quantum Hall edge states, since they exist not in the whole edge BZ, but in its finite subset, whose size is determined by the momentum-space separation of the Dirac nodes. Finally, we find that the Weyl semimetal has a finite DC conductivity at zero temperature, but Drude weight vanishing as $T^2$,  and 
is thus an interesting metallic state, characterized by a nonzero anomalous Hall conductivity and topologically-protected edge states. 
Interesting open questions include the influence of Coulomb interactions on the properties of Weyl semimetals, in particular their transport properties.

\begin{acknowledgments}
We acknowledge useful discussions with Matthew P.A. Fisher, Ying Ran and Cenke Xu. Financial support was provided by the NSERC of Canada and a University of Waterloo start-up grant (AAB), by NSF grants DMR-0804564 and PHY05-51164 (LB), and by
the Army Research Office through MURI grant No. W911-NF-09-1-0398 (LB). AAB gratefully acknowledges the hospitality of KITP, where 
this work was done.

\end{acknowledgments}

\end{document}